\documentclass[twocolumn,showpacs,preprintnumbers,superscriptaddress,
pra,floatfix]{revtex4}
\usepackage[brazil]{babel}
\usepackage{graphicx}
\usepackage{dcolumn}
\usepackage{bm}
\usepackage{natbib}
\usepackage[latin1]{inputenc}

\newcommand{\ket}[1]{\left| #1 \right>}

\newcommand{\brahket}[3]{\left< #1 \left| #2 \right| #3 \right>}

\begin{document}

\title{Quantum Entanglement and fixed point Hopf bifurcation}
\author{M. C. Nemes} \thanks{ E--mail: carolina@fisica.ufmg.br }
\affiliation{Departamento de Física, ICEX,
 Universidade Federal de Minas Gerais, C.P. 702,
 30161-970, Belo Horizonte, M.G., Brazil}
\author{K. Furuya} \thanks{ E--mail: furuya@ifi.unicamp.br},
\affiliation{Instituto de Física `Gleb Wataghin',
 Universidade Estadual de Campinas,\\
 C.P. 6165, 13083-970, Campinas, S.P., Brazil}
\author{G. Q. Pellegrino}
\affiliation{Departamento de Matemática,
Instituto de Geociências e Ciências Exatas, Universidade Estadual Paulista,\\
 13500-230, Rio Claro, S.P., Brazil}
\author{A. C. Oliveira}
\affiliation{Departamento de Ciências Exatas e Tecnológicas,
Universidade Estadual de Santa Cruz,\\
45662-000, Ilhéus, BA, Brazil}
\author{Maurício Reis}
\affiliation{Departamento de Física, ICEX,
 Universidade Federal de Minas Gerais, C.P. 702,
 30161-970, Belo Horizonte, M.G., Brazil}
\author{L. Sanz}
\affiliation{ Departamento de F\'{\i}sica, Universidade Federal
de S\~ao Carlos, 13565-905, S\~ao Carlos, SP, Brazil}
\begin{abstract}
We present the qualitative differences in the phase transitions of the
mono-mode Dicke model in its integrable and chaotic versions. We show
that a first order phase transition occurs in the integrable case whereas
a second order in the chaotic one. This difference is also reflected in
the classical limit: for the integrable case the stable fixed point in
phase space suffers a bifurcation of Hopf type whereas for the second one
a pitchfork type bifurcation has been reported.
\end{abstract}
\maketitle
\section{Introduction}
  Entanglement is a fundamental characteristic of quantum systems which
has lately provided for impressive progress in several areas such as
quantum information \cite{information}, quantum cryptography \cite{crypto}
and teleportation \cite{teleport}. In this context, several measures of
this property have been proposed, such as entropy-like quantities
\cite{entropy}, negativity \cite{negativ}, concurrence \cite{concurrence}
and so forth.
In particular, one expects strong correlations to be present in quantum phase
transitions (QPT) \cite{qpt}. Therefore, in this context it seems particularly
 interesting to study the interplay between quantum correlations and the
appearance of a QPT.
Since the quantification of the degree of entanglement (or correlations) in
a system is not unique, in recent years several examples of specific
systems using different measures have been put forth. More specifically,
ground state correlations have been investigated as a function of a coupling
parameter. In references
\cite{schneider02,osborne02,Gvidal03,costi03,Jvidal04a,Jvidal04b,lambert04,buzek05} the ground state two atom concurrence has been studied and shown
to exhibit a maximum at the transition point, in the context of some spin-spin
or spin-boson models \cite{note1}.
As pointed out recently by Hines-McKenzie-Milburn (HMM) \cite{hines05}, in the
classical regime corresponding to the qualitative change in the quantum ground
state it corresponds a change in minimum energy stable fixed point
in phase space. As the parameter value of the Hamiltonian is changed, a
quantum instability of the ground state is followed (at $\lambda=\lambda_c$),
and a qualitative change in phase space structure of the system.
In their work, the pitchfork (emergence of two new stable fixed point) type
bifurcation has been focused and associated to a spike of the entanglement
in the mean field limit ($N \rightarrow \infty$).

 The purpose of the present work is firstly to add to these previous
contributions a study of an integrable version of Dicke's model
\cite{dicke,tavis}, and to show that the phase transition in this case
is radically different from the one in the non-integrable situation as
revealed by the linear entropy (adopted here as a measure of quantum
correlations)\cite{note2}.
Reference \cite{hines05} proposes, based on specific model studies how
the entanglement in a nonlinear bipartite system can be associated with
a fixed point bifurcation in the classical dynamics. This conjecture thus
contemplates pitchfork (one-to-two) bifurcation. Our results here suggest
that first order transitions can present a different type of bifurcation.
We show that in the integrable version of the model the abrupt
change in entanglement content of the ground state (GS) is associated to a
Hopf type bifurcation (one-to- infinitely many degenerate fixed points).
In fact this also occurs for the Jahn-Teller \cite{hines04} and dimer
\cite{dimer} models.
So, we suggest that when the QPT is first order the corresponding classical
instability is of a Hopf type bifurcation. This idea is supported by the
analysis of the GS Wigner function in the integrable and non-integrable
situations. The maxima of these functions exactly follow the classical fixed
points.
 \section{ the model and the behavior of the entropy}
   The quantum Hamiltonian we use is written in the form
\begin{eqnarray}
\hat{H} =\hat{H}_o+ \hat{H}_G+\hat{H}_G^{'}
\label{Hdicke}
\end{eqnarray}
with
\begin{eqnarray}
\hat{H}_o & = & \hbar\omega\hat{a}^{\dagger}\hat{a}+\hbar\epsilon
\hat{J}_z \nonumber\\
\hat{H}_G & = & \frac{G}{\sqrt{2J}}\left(\hat{a}\hat{J}_{+}+\hat{a}^{\dagger}
\hat{J}_{-} \right) \nonumber\\
\hat{H}_{G^{'}} & = & \frac{G^{\prime}}{\sqrt{2J}}
\left(\hat{a}^{\dagger}\hat{J}_++\hat{a}\hat{J}_-\right).
\end{eqnarray}
where we consider in all numerical calculations shown $\omega=\epsilon=1$
(resonant case).
When $G=G'$ we recover the usual single mode Dicke Hamiltonian.
The advantage of working with Hamiltonian (\ref{Hdicke}) is that the
integrable regime can be obtained just by setting $G' = 0$ (or $G = 0$),
and it becomes easier to explore various mixed regimes. The Super-radiant
phase transition is present in all these situations at $G+G'=\epsilon$
\cite{superradiance}.
\begin{figure}[ht]
\includegraphics[scale=0.35,angle=0]{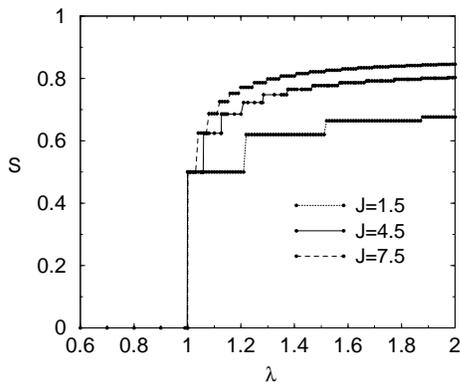}
\hspace{6.0cm}
\vspace{-1.0cm}
\caption{Linear entropy $S$ versus $\lambda=G/\epsilon$ for parameter values
$J=1.5,4.5,7.5$, $\hbar=1$,$\omega=\epsilon=1$ : integrable case ($G'=0$),
with a clear transition at $\lambda_c=1$.}
\label{entropy}
\end{figure}

The non-integrable GS features of this model have been studied in ref.
\cite{lambert04} using the concurrence as a measure of GS correlations. It
is shown that concurrence attains a maximum when $\lambda=\lambda_m(N)$.
This maximum is a function of the number of atoms and tends to the critical
value $\lambda_c=\lambda_m(\infty)$ as $N \rightarrow \infty$. In the present
work we use the linear entropy of the atomic subsystem as a measure of the
quantum correlations.
We observe the same qualitative behavior which is therefore not shown here.
However, for the integrable situation $G' = 0$ shown in Fig.\ref{entropy},
the atomic linear entropy (ALE) is plotted as a function of
$\lambda=G/\epsilon$ for $N=2J=3,9,15$ atoms. The ALE presents an abrupt jump
from zero to $0.5$ at $\lambda_c=G/\epsilon=1$ for {\sl all} values of $J$.
Moreover, the entropy has a monotonic increase in steps of finite size:
for example, before the first jump ($\lambda<\lambda_c$) only {\sl one} basis
state contributes to the ground state.
It is a well known fact that below $\lambda=\lambda_c=1$ the ground state
of the integrable version of the model is an isolated basis state
$|n=0,m=-J \rangle$. It is not connected to any other basis state via
the above interaction for $G'=0$ \cite{tavis}, until the
interaction term of the Hamiltonian becomes dominant and the ground state
becomes a state of the form  $\frac{1}{\sqrt{2}}\left(|1,-J  \rangle
+|0,-J+1 \rangle \right)$, which is orthogonal to the previous fundamental
state. The larger the value of $\lambda$, the smaller the steps
become, and finally a plateau is attained for large values of the coupling.
The plateau values of the entropy increases with increasing value of $J$,
indicating a larger number of basis states participating in the entanglement
of the two subsystems.
 At the limit $J>>1$ we expect the height of the steps to remain of the same
size, but the width to become shorter and shorter, whereas the plateau tends
to $1$.
In this case, the classical analog of this behavior stems from the fact
that in the classical limit the equilibrium points present bifurcations
at the same parameter values as in the quantum case. The bifurcations as
shown in \cite{qoptics91}, in both - integrable and non-integrable - cases
are also very distinct. The classical limit is discussed below.

\section{\sl The classical analog and bifurcation of equilibria}

The classical Hamiltonian corresponding to eq. (\ref{Hdicke})
can be obtained by a standard procedure using the coherent states
of spin and boson \cite{klauder85}:
\begin{equation}
\ket{w} = \left( {1+w\bar w} \right)^{-J}e^{wJ_+} \ket{J, -J} \label{SPIN}
\end{equation}
\begin{equation}
\ket{v} = e^{-v\bar v/2}e^{vb^+} \ket{0} \label{FIELD}
\end{equation}
with
\begin{equation}
w = {{p_1+iq_1} \over {\sqrt {4J-\left( {p_1^2+q_1^2} \right)}}},
\label{pq1}
\end{equation}
\begin{equation}
v = {1 \over {\sqrt 2}}\left( {p_2+iq_2} \right).
\label{pq2}
\end{equation}
The classical Hamiltonian is defined as  $\brahket{wv}{H}{wv}$ \cite{annals},
and in this case there results a nonlinearly  coupled two degrees of freedom
Hamiltonian function in terms of the parametrization (\ref{pq1}) and
(\ref{pq2}):
\begin{eqnarray}
{\cal H}(q_1,p_1,q_2,p_2) &=&
{{\omega} \over 2}\left( {p_2^2+q_2^2} \right)+
{\varepsilon  \over 2}\left( {p_1^2+q_1^2} \right) - \varepsilon J +
\nonumber\\
&&{{\sqrt {4J-\left( {p_1^2+q_1^2} \right)}}
\over {\sqrt {4J}}}\left( {G_+p_1p_2+G_-q_1q_2} \right),\nonumber\\
\end{eqnarray}
where $G_{\pm} = G \pm G'$. The integrable situation corresponds to $G'=0$ and
$G_{\pm} = G$.\\

  The stable fixed points corresponding to the equilibrium positions in
phase space are defined via Hamilton's equations:
\begin{eqnarray}
\label{eqmov}
\dot{q_1} = -\epsilon p_1 - \frac{G_+ p_2}{\sqrt{2J}}(2J-H_1)^{1/2}
+\nonumber\\
+ \frac{p_1}{\sqrt{2J}(2J-H_1)^{1/2}}(G_+ p_1p_2 +G_- q_1q_2)=0\\
\dot{p_1} = -\epsilon q_1 + \frac{G_- q_2}{\sqrt{2J}}(2J-H_1)^{1/2}
+\nonumber\\+ \frac{q_1}{\sqrt{2J}(2J-H_1)^{1/2}}(G_+ p_1p_2+G_-q_1q_2)=0\\
\dot{q_2} = -\omega p_2  - \frac{G_+ p_1}{\sqrt{2J}}(2J-H_1)^{1/2}=0\\
\dot{p_2} = \omega q_2  + \frac{G_- q_1}{\sqrt{2J}}(2J-H_1)^{1/2}=0
\end{eqnarray}
where $H_1=\frac{q_1^2+p_1^2}{2}$.\\

In ref. \cite{qoptics91} it has been
shown that there are two solutions for:\\
A. {\sl Integrable case}: \\
i) trivial solution (stable for $G/\epsilon < 1$) :
\begin{eqnarray}
 q_1=p_1=q_2=p_2=0 \textrm{, (the origin)}
\label{null}
\end{eqnarray}
ii) non-trivial solution (stable for $G/\epsilon > 1$):
\begin{eqnarray}
 R_1^2=q_1^2+p_1^2= 2J \left( 1-\frac{\epsilon \omega}{G^2}\right)
\label{radiia}
\end{eqnarray}
\begin{eqnarray}
 R_2^2=q_2^2+p_2^2= J \left( \frac{G^4-\epsilon^2\omega^2}{G^2\omega^2}\right)
\label{radiif}
\end{eqnarray}
The above equations (\ref{null},\ref{radiia},\ref{radiif})  show in each
phase space a  {\sl Hopf} bifurcation of classical equilibria:
we have a single point for the minimum energy $E_{GS}$ before the transition
and an infinitely degenerate state represented in phase space by a circle.

B. {\sl Chaotic regime}: the equilibrium point below
$G_+/\epsilon=1$ bifurcates to two points \cite{qoptics91}:\\
i) trivial solution, stable for $G_+/\epsilon < 1$ (and therefore
$G_-/\epsilon < 1$ ):
\begin{eqnarray}
 q_1=p_1=q_2=p_2=0 \textrm{, (the origin)}
\label{null2}
\end{eqnarray}
ii) non-trivial solution I, stable for $G_+/\epsilon > 1$ and
$G_-/\epsilon < 1$ corresponding to the {\sl pitchfork} bifurcation in each
phase space:
\begin{eqnarray}
 q_1=q_2=0 \textrm{,   }
 p_1=\pm  \left(\frac{2J(G_+^2-\epsilon \omega)}{G_+^2}\right)^{1/2}
\textrm{  ,}\nonumber\\
 p_2=\mp  \left(\frac{J(G_+^4-\epsilon^2 \omega^2)}{\omega^2 G_+^2}
\right)^{1/2}.
\label{psol}
\end{eqnarray}
iii) non-trivial solution II, stable for $G_+/\epsilon > 1$ and
$G_-/\epsilon > 1$:
\begin{eqnarray}
 p_1=p_2=0 \textrm{,   }
 q_1=\pm  \left(\frac{2J(G_-^2-\epsilon \omega)}{G_-^2}\right)^{1/2}
\textrm{  ,}\nonumber\\
 q_2=\mp  \left(\frac{J(G_-^4-\epsilon^2 \omega^2)}{\omega^2 G_-^2}
\right)^{1/2}.
\label{qsol}
\end{eqnarray}
\section{Atomic Wigner Functions}
The maxima of the atomic Wigner distributions of the quantum ground state
displays amazing similarity with the classical bifurcations:\\

1) for $\lambda < \lambda_c$ the atomic Wigner function (AWF) in both cases
are a Gaussian-like state centered at the classical equilibrium position (the
origin - not shown).\\
\begin{figure}[ht]
\includegraphics[scale=0.5,angle=0]{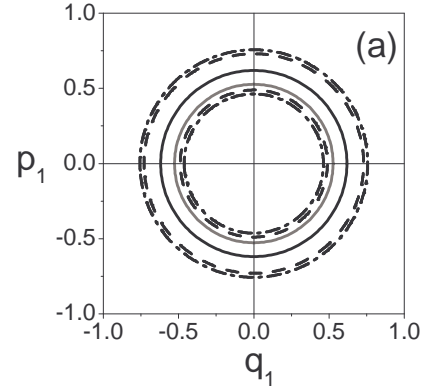}
\includegraphics[scale=0.3,angle=-90]{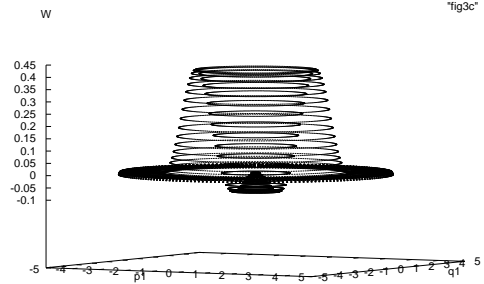}
\caption{(a) contour plots at the height of $50\%$ of the Integrable Atomic
Wigner functions of the ground state for: $\omega=\epsilon=1$,
$\lambda=G/\epsilon=1.5$, $G'=0$ and scaled variables $(q_1/\sqrt{4J},p_1/
\sqrt{4J})$. Dot-dashed line -- $J=4.5$, dashed line  -- $J= 10.5$. The
{\sl black continuous} line -- maximal region of the AWF, and the {\sl gray
continuous} line corresponds to the projection of the minimum energy region
 of classical fixed points as given by Eq.(\ref{radiia}).
(b) AWF showing the negative part for $J= 10.5$.}
\label{wigI}
\end{figure}
2) for $\lambda > \lambda_c$, we separate the two cases:\\
(a) {\sl Hopf bifurcation}: for the integrable case, there is only one
non-trivial
regime ($\lambda=G/\epsilon > \lambda_c$), and we show in Fig.\ref{wigI}a
a contour plot of the  AWF on the scaled atomic phase space $(q_1/\sqrt{4J},
p_1/\sqrt{4J})$. These figures correspond  to the ground state of the
integrable system for $J=4.5$ and $10.5$ respectively at the same value of
$\lambda = G/\epsilon = 1.5 > \lambda_c$. Note that the projection of the
region covered by the maxima is an annular region for both cases, and is
shown in black continuous line the circle corresponding to the maxima of
the AWF. In gray continuous line we show the region corresponding to the
bifurcated classical stable fixed points. This shows that although
qualitatively the mean field predicts correctly the change in the GS, the
result does not give the position of the radius correctly. It is well known
that mean field type calculations predict very well the energy (as shown
for this model in \cite{qoptics91}), but the states are not so well described.
In Fig.\ref{wigI}b a 3-dimensional plot is shown, with its negative part
clearly visualizable for the case $J=10.5$.\\

\begin{figure}[ht]
\includegraphics[scale=0.35,angle=-90]{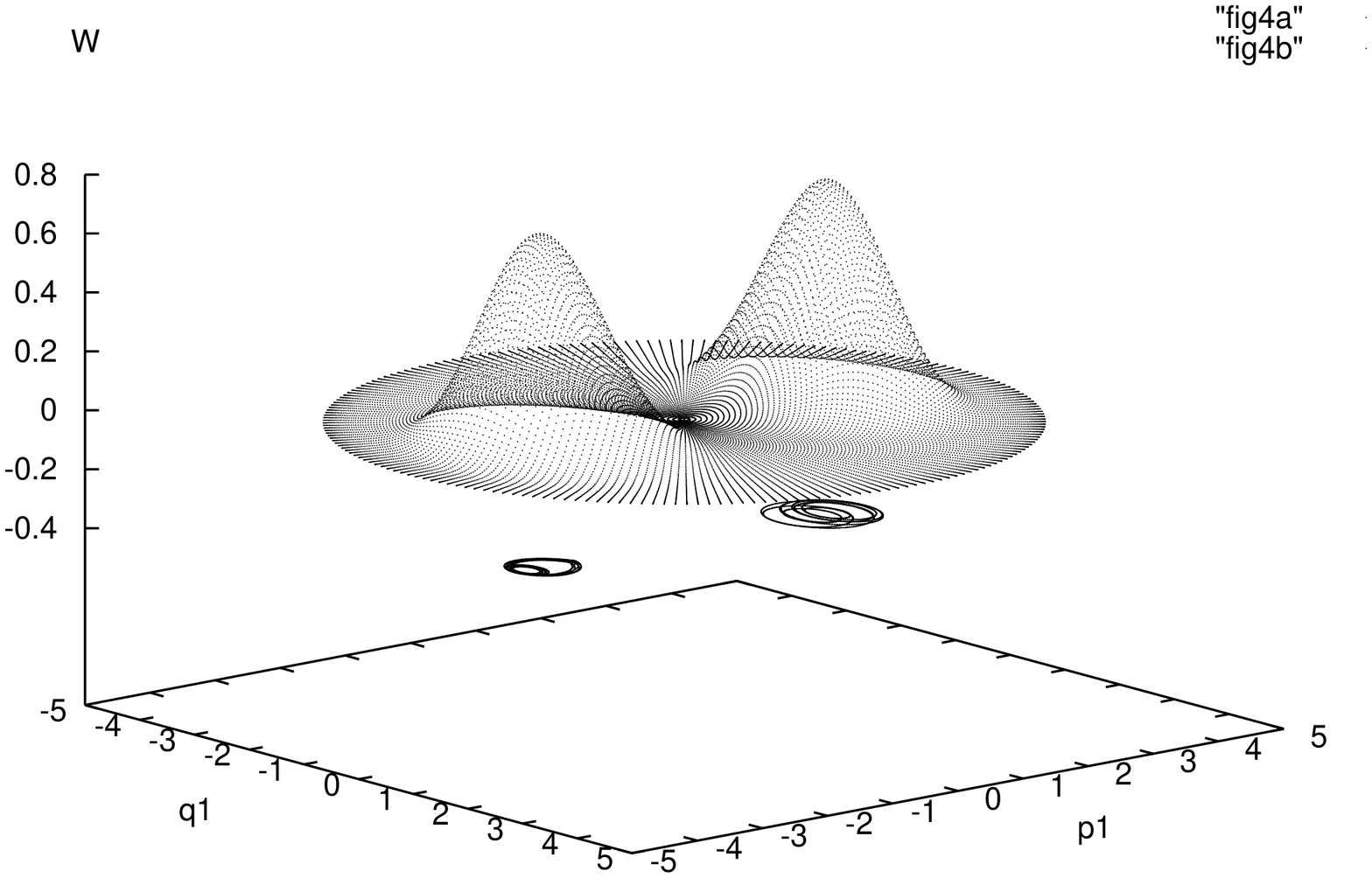}
\includegraphics[scale=0.35,angle=-90]{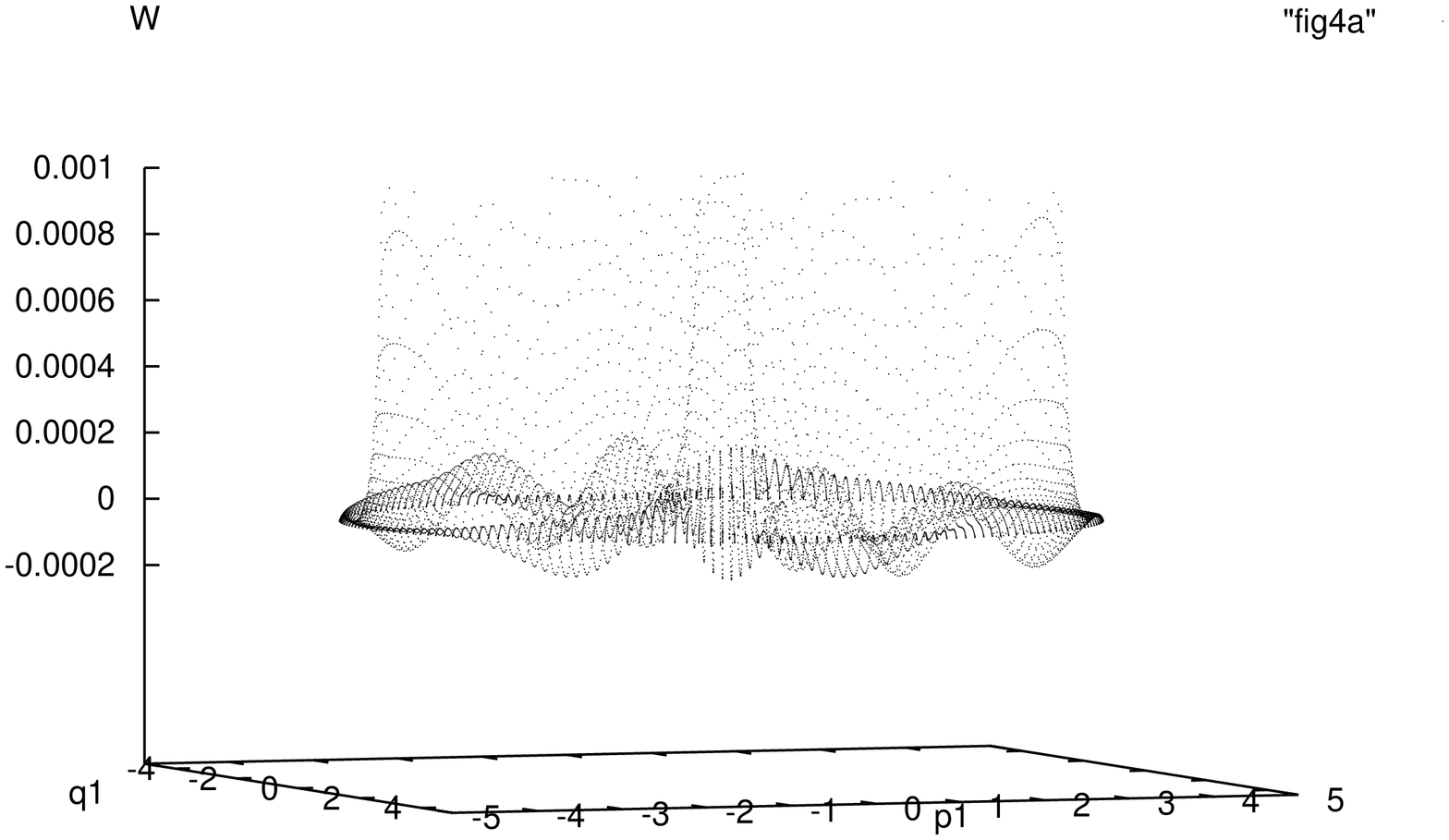}
\caption{(a) Non-Integrable Atomic Wigner function of the ground state after
the phase transition and the classical trajectories for $G/\epsilon=G'/
\epsilon=0.75$, $J=4.5$ and $E=-5.5$, projected onto the atomic phase space
with scaled variables $(q_1/\sqrt{4J},p_1/\sqrt{4J})$; (b) another view of
the same AWF showing the negative part.}
\label{wigNI}
\end{figure}

(b) {\sl pitchfork bifurcation}: for the non-integrable case, one regime is
shown ($\lambda_+ = G_+/\epsilon > \lambda_c$ and $\lambda_-=G_-/\epsilon <
\lambda_c$) the non-trivial solution I is shown in Fig.\ref{wigNI}. Again,
the maxima of the AWF clearly splits into a pair of peaks, connecting to the
pitchfork type bifurcation shown in the classical case, and discussed in
\cite{hines05}.\\
In order to estimate the number of basis states participating in the
entanglement for different values of $N$, we calculated the phase space
area (in units of $\hbar$) at the  height of $50\%$ of the AWF shown in
Fig.\ref{wigI}a for $N=10$ and $N=21$. It becomes clear that it increases
with $N$: $ a_{10}= \frac{ A(J=4.5)}{10\hbar})=3.58$ and
$ a_{21}= \frac{ A(J=10.5)}{21\hbar})=5.32$, thus $ a_{21} > a_{10}$.\\

 As for the non-integrable regime, Fig.\ref{wigNI}a shows a very distinct
AWF with two bumps centered at the bifurcated classical equilibrium points.
We also plot the classical trajectories projected onto the atomic phase
space for an energy slightly above the value of the GS. Note again the close
analogy: the trajectories are concentrated in two distinct regions whose
center correspond to the center of the two peaks in the quantum AWF.
Fig.\ref{wigNI}b shows quantum interference effects, in the negative part
of the AWF, but it is very small, not allowing to confirm the suspicion
of ref. \cite{emary03} that this GS is a Shr\"odinger-cat like state.
It is also worth noting that there are quantitative as well as qualitative
differences in what concerns the negative parts of the Wigner
functions. In the integrable situation, its negative part is
distributed in narrower region of phase space as compared with the
non-integrable one, where negative contributions seem to be spread over a
larger area. However, from the quantitative point of view, the integrable
AWF exhibits a deeper negative contribution \cite{note3}.

\section{Conclusions}

In summary we have shown that the phase transition in
Dicke's model can be qualitatively different in the {\sl integrable regime}
and in the non-integrable one, the last case already discussed by other
authors, although using the concurrence as a measure of entanglement. The
linear entropy produces the same qualitative results.

   The relationship between entanglement in the ground state at
transition point, and the phase space bifurcations in the classical limit
has already been noted in the literature. We have found here an example
of how the type of the quantum phase transition may qualitatively alter the
corresponding classical instability. In the case of the Dicke model,
the integrable and non-integrable situations are shown to be qualitatively
very different.
In quantum terms the GS exhibits a first order QPT reflected
in the classical limit as a Hopf bifurcation which is very different from
the already discussed non-integrable situation (second order QPT and pitchfork
bifurcation). The close  quantum and classical analogy is also strongly
reflected in the corresponding GS atomic Wigner functions.
\begin{acknowledgments}
It is a pleasure to acknowledge W. D. Heiss for encouraging the present work.
We thank financial support from CAPES and CNPq (Conselho Nacional de
Pesquisa, Brazil) and FAPESP.
\end{acknowledgments}


\begin{thebibliography}{99}
\bibitem{information}  M.A. Nielsen and I.L. Chuang, ``Quantum Computation
and Quantum Information'' (Cambridge University Press, Cambridge, 2000).

\bibitem{crypto} C.H. Bennett, G. Brassard, C. Crepeau, U.M. Maurer,
IEEE Trans. Inf. Theor.  {\bf 41} (6): 1915 (1995);
N. Gisin, G.G. Ribordy, W. Tittel , H. Zbinden, Rev Mod. Phys. {\bf 74} (1):
145 (2002).

\bibitem{teleport} C.H. Bennett, {\sl et al.}, Phys. Rev. Lett. 70 (13):
1895 (1993); D. Bouwmeester, {\sl et al.}, Nature {\bf  390} (6660): 575
(1997); for a review, see: Alber G, {\sl  Quantum Information }, Springer
Tracts in Modern Physics {\bf 173}: 1 (2001).

\bibitem{entropy} V. Vedral, Rev. Mod. Phys. {\bf 74}, 197 (2002)

\bibitem{negativ} K. Zyczkowski, P. Horodecki, A. Sanpera, M. Lewenstein,
Phys. Rev. A {\bf 58}, 883 (1998).

\bibitem{concurrence}  W.K. Wootters, Phys. Rev. Lett. {\bf 80}, 2245 (1998).

\bibitem{qpt}  S. Sachdev, {\sl  Quantum Phase Transitions} (Cambridge
University Press, Cambridge, 2000).

\bibitem{schneider02} S. Schneider and G.J. Milburn,
Phys. Rev. A {\bf 65} , 042107 (2002).

\bibitem{osborne02} T.J. Osborne, and M.A. Nielsen,
Phys. Rev. A {\bf 66} , 032110 (2002).

\bibitem{Gvidal03} G. Vidal, J.I. Latorre, E. Rico,A. Kitaev,
Phys. Rev. Lett. {\bf 90}, 0227902 (2003).
%
\bibitem{costi03} T.A. Costi and R.H. McKenzie,
Phys. Rev. A {\bf 68} , 034301 (2003).

\bibitem{Jvidal04a} J. Vidal, R. Mosseri, and J. Dukelsky,
Phys. Rev. A {\bf 69} , 054101 (2004).

\bibitem{Jvidal04b} J. Vidal, G. Palacios, and R. Mosseri,
Phys. Rev. A {\bf 69} , 022107 (2004).

\bibitem{lambert04} N. Lambert, C. Emary, and T. Brandes,
Phys. Rev. Lett. {\bf 92}, 073602 (2004); N. Lambert, C. Emary, and
T. Brandes,  Phys. Rev. A 71, 053804 (2005).

\bibitem{buzek05} V. Buzek, M. Orszag and M. Rosko,
Phys. Rev. Lett. {\bf 94}, 163601 (2005).

\bibitem{note1} In ref. \cite{yang05} a specific spin model with three particle interactions is presented as a counterexample when a discontinuity in the
first derivative of the G.S. concurrence appears also in the absence of a QPT.

\bibitem{hines05} A.P. Hines, R.H. McKenzie, and G.J. Milburn,
Phys. Rev. A {\bf 71} , 042303 (2005).

\bibitem{dicke}  R.H. Dicke,
Phys. Rev. {\bf 93}, 99 (1954).

\bibitem{tavis} M. Tavis and F.W. Cummings,
Phys. Rev. 170 (2): 379 (1968).

\bibitem{note2} after the completion of this work we become aware
of the work \cite{buzek05}, where also this sequence of instabilities
in the GS of the integrable model has been discussed using the concurrence
of bipartite systems.

\bibitem{hines04} A.P. Hines, C.M. Dawson, R.H. McKenzie,G.J. Milburn,
Phys. Rev. A {\bf 70} , 022303 (2004).

\bibitem{dimer} X.-W. Hou, J.-H. Chen, and B. Hu,
Phys. Rev. A {\bf 71} , 034302 (2005).

\bibitem{superradiance}{\sl Cooperative Effects in Matter and Radiation},
edited by C.M. Bowden, D.W. Howgate, and H.R. Robl (Plenum, New York, 1977).

\bibitem{qoptics91} M.A.M. de Aguiar, K. Furuya and M.C. Nemes,
 Quant. Opt. {\bf 3} (1991) 305.

\bibitem{klauder85} J. R. Klauder and B.-S. Skagerstam, {\em Coherent States:
Applications in Physics and Mathematical Physics} (World Scientific, Singapore,
1985).

\bibitem{annals} M.A.M. de Aguiar, K. Furuya, C.H. Lewenkopf, and M.C. Nemes,
 Ann. Phys. {\bf 216} (2) (1992) 291.

\bibitem{manybody} J.I. Latorre {\sl et al.},
cond-mat/0409611.

\bibitem{emary03} C. Emary and T. Brandes,
Phys. Rev. Lett. {\bf 90}, 044101 (2003);  C. Emary and T. Brandes,
Phys. Rev. E {\bf 67} , 066203 (2003).

\bibitem{note3} Concerning the chaotic atomic Wigner functions of
the non-integrable version of the Dicke model, a more detailed study
will be shown elsewhere.

\bibitem{yang05} M.-F. Yang, Phys. Rev. A {\bf 71} , 030302(R) (2005).

\bibitem{glendinning}P. Glendinning, {\sl Instability and Chaos: An Introduction to
the Theory of Nonlinear Differential Equations}, (Cambridge University Press,
Cambridge, U.K., 1994).
\end{thebibliography}
\end{document}